\documentclass[letterpaper, 10pt, conference]{ieeeconf}
\IEEEoverridecommandlockouts

\usepackage{floatrow}
\usepackage{afterpage}
\usepackage{cite}
\usepackage{algorithm} 
\usepackage{algorithmic}  
\usepackage[algo2e]{algorithm2e} 
\usepackage{amsmath,amssymb,amsfonts}
\usepackage{algorithm}
\usepackage{graphicx}
\usepackage{textcomp}
\def\BibTeX{{\rm B\kern-.05em{\sc i\kern-.025em b}\kern-.08em
    T\kern-.1667em\lower.7ex\hbox{E}\kern-.125emX}}
    
    \usepackage{lineno}
\usepackage{lettrine}

\usepackage{moreverb}

\usepackage{graphicx}
\usepackage{subcaption}
\usepackage{float}
\usepackage{amsmath}
\usepackage{array}
\usepackage{multirow}
\graphicspath{ {Figures/} }

\usepackage{scalerel}
\usepackage{tikz}
\usetikzlibrary{svg.path}

\definecolor{orcidlogocol}{HTML}{A6CE39}
\tikzset{
  orcidlogo/.pic={
    \fill[orcidlogocol] svg{M256,128c0,70.7-57.3,128-128,128C57.3,256,0,198.7,0,128C0,57.3,57.3,0,128,0C198.7,0,256,57.3,256,128z};
    \fill[white] svg{M86.3,186.2H70.9V79.1h15.4v48.4V186.2z}
                 svg{M108.9,79.1h41.6c39.6,0,57,28.3,57,53.6c0,27.5-21.5,53.6-56.8,53.6h-41.8V79.1z M124.3,172.4h24.5c34.9,0,42.9-26.5,42.9-39.7c0-21.5-13.7-39.7-43.7-39.7h-23.7V172.4z}
                 svg{M88.7,56.8c0,5.5-4.5,10.1-10.1,10.1c-5.6,0-10.1-4.6-10.1-10.1c0-5.6,4.5-10.1,10.1-10.1C84.2,46.7,88.7,51.3,88.7,56.8z};
  }
}

\newcommand\orcidicon[1]{\href{https://orcid.org/#1}{\mbox{\scalerel*{
\begin{tikzpicture}[yscale=-1,transform shape]
\pic{orcidlogo};
\end{tikzpicture}
}{|}}}}

\usepackage{tikz}
\usepackage{qrcode}
\usepackage[printonlyused]{acronym}

\usepackage[draft]{hyperref}
\floatstyle{plaintop}
\restylefloat{table}

\begin{document}
\title{\LARGE \bf
User-to-Vehicle Interaction in Smart Mobility: The GO-DRiVeS Autonomous Ride-Sharing Application}

\author{Hana E. Elmalah$^{1,2\orcidicon{0009-0004-0617-6776}\,}$ and Catherine~M.~Elias$^{1,2\orcidicon{0000-0002-1444-9816}\,}$,~\IEEEmembership{Member,~IEEE,}%
\thanks{*This work was not supported by any organization}
\thanks{$^{1}$C-DRiVeS Lab: Cognitive Driving Research in Vehicular Systems, Cairo, Egypt
{\tt\small cdrives.researchlab@gmail.com}}%
\thanks{$^{2}$Computer Science and Engineering Department - Faculty of Media Engineering and Technology - German University in Cairo, Egypt}%
\thanks{{\tt\scriptsize hana.elmalah@student.guc.edu.eg, catherine.elias@ieee.org}}%
}

\markboth{Journal of \LaTeX\ Class Files,~Vol.~14, No.~8, August~2015}%
{author1 \MakeLowercase{\textit{et al.}}:title here}
%



\maketitle
\begin{abstract}
This paper introduces the  GO-DRiVeS application, an on demand ride sharing and requesting mobile application tailored specifically to save long walks and challenges which are time consuming and tiring especially during hot days or when carrying heavy items, faced by university students and staff. The GO-DRiVeS application was developed following the Agile methodology for its flexibility. In addition to, using the mobile application system architecture and client-server architecture. GO-DRiVeS was implemented using React Native (Expo) for the frontend, Node.js and Express for the backend, and MongoDB as the database; based on a detailed analyses to the existing transportation application, comparing their frameworks and identifying their essential functionalities. GO-DRiVeS supports core features like user registration, ride requesting and real-time tracking.In addition to handling multiple requests at the same time in a first come first serve manner. The application was developed based on these features, and the results were conducted in the form of multiple experiments that demonstrated stable behavior in handling the requests, as presented in the Methodology and Results chapters.
\end{abstract}

\begin{keywords}
  Ride-Sharing in Campus, University Students , Android , Mobile Application , GPS Navigation
\end{keywords}
\IEEEpeerreviewmaketitle
\section{Introduction and Related Work}\label{sec1}
In former times, when people were asked how they thought their daily lives would be like in the future, almost everyone agreed that technology will play a huge role in our lives, as it will be in nearly every day activity, as almost the majority of people back then believed in having a future filled with Autonomous vehicles, according to a survey by \cite{kyriakidis2015}.

Actually, their predictions of the future being filled with Autonomous vehicles is in fact happening and will increase more and more in the future according to the statistics done on the number of autonomous vehicles expected to be reached until 2032 \cite{marketus2025}.

Nowadays the usage of autonomous vehicles reached the transportation like Waymo, which is a self driven car that is used like taxis but they are not like any other taxi as it is autonomous, so how will people use and interact with it. That is where ride hailing mobile applications come in handy and this is exactly what Waymo used to facilitate the interactions between the users and their autonomous vehicles. 

After reviewing several existing ride sharing applications for university students, it was found that there are multiple different tools and technologies used to create these apps. In this section a comparison will be made between each of the Frontend, Backend, Database, and Mapping tools used in those applications.

The literature review related to this work is mainly divided into three parts: the fronend, the backend, and the database.

In terms of the frontend state of the art, and after previewing the work presented in \cite{p6,p3,p5,p7,p12,p13,p16,p17,p1,p2,p4,p8} it was found that react native and Android Studio are the most used frameworks, but actually according to statistics \cite{jetbrains2025crossplatform,pangea2025frameworks} done on the most used frontend framework for mobile application development in 2025; it was found that Flutter and React native are the most used; as they provide a combination of good performance and developing experience.

Flutter is popular these days and  is regarded to be the first option for frontend development to some people due to its ability to operate in multi-operating systems, and for its rich and lovely customizable User Interface.
On the contrary, React Native is preferred by others , in view of the fact that it is a JavaScript based framework which is recognized by most developers, in addition to having the option to make the code shareable between web
and mobile platforms via Expo, furthermore  having a variety of libraries, plugins, and
community tools.

on the other hand the backend state of the art is presented in \cite{p1,p3,p4,p7,p8,p9,p17,p2,p6,p12,p13,p16}, it was found that Node.js and Firebase are the most used backend frameworks, and in fact they are; according to statistics \cite{konstant_backend_frameworks,ethan2025backend} done on the most used backend frameworks for mobile applications in 2025.

Node.JS is considered to be the primary choice for many  developers for the fact that it is a JavaScript based framework, which is recognized by most, in addition to having an event-driven architecture that manages multiple requests at once. It also has NPM libraries that cover almost all backend needs.

Firebase is also preferred by others due to its server-like backend and  real-time database, besides its simple and easy iOS and Android integration.
Finaly the database state of the art and in mobile app development, there are two primary database types used SQL and NoSQL, and based on the structure and requirements of the applications the type that should be used is determined. After reviewing these work presented in \cite{p3,p17,p1,p2,p4,p5,p6,p7,p12,p13,p16,p9,p8}, it was observed that MongoDB and Firebase are the most common database framework for the NoSQL type used. Furthermore, these two frameworks were regarded the most popular NoSQL databases for mobile applications in 2024 due to their features, per the statistics \cite{risingwave2024nosql} on the subject.  Whereas,  PostgreSQL and MySQL were noticed to be the most common SQL database frameworks, and  according to statistics \cite{zucisystems2025} done on the most used SQL frameworks in mobile development these two frameworks were the top ones used.

Although several ride-hailing and mobility applications exist such as Uber, Lyft and Bolt, most of them follow the multi-application structure, which is to have an application specific to the user and another different one for the driver. Moreover, there is an absence and no documentation for ride-sharing applications on autonomous cars specifically for university campuses in Egypt. Thus, there is a gap in the development of a unified,scalable and locally adapted ride sharing application on autonomous cars that are specially tailored to Egyptian university campuses.

In order to achieve the objective, the GO-DRiVeS application will be developed using: Frontend: React Native (Expo), Backend: Node.js and Express, Database: MongoDB, Communication: Socket.IO, Map: Expo map (Google maps), Routing: OpenRouteService, Tracking: Expo location.

This research aims to:
\begin{itemize}
    \item Develop a scalable, real-time, unified ride sharing application specifically for university campuses in Egypt.
    \item Allow users to request rides in real time and allow drivers to manage ride requests through communications.
    \item Implement routes that are optimized to improve trip's efficiency.
    \item Implement real-time GPS tracking for cars.
\end{itemize}
\section{Methodology}\label{sec2}
This section provides an overview of the methodology followed in the design, development, and implementation phases of the GO-DRiVeS application. However, a more thorough introduction to the app should come first. GO-DRiVeS is a ride requesting and sharing application tailored specifically for campus transportation on campus autonomous vehicles. Admin, Cars (drivers), and Users (faculty or students) are its three stackholders. 

The Agile project management methodology was selected to be used in the GO-DRiVeS application for its iterative development; as it divides the project into
 small, manageable chunks which are called sprints,hence having the ability to develop a core functionality and then test it without having to finish all the other functionalities. 
 
\begin{figure}[h]
\centering
\includegraphics[width=0.9\textwidth]{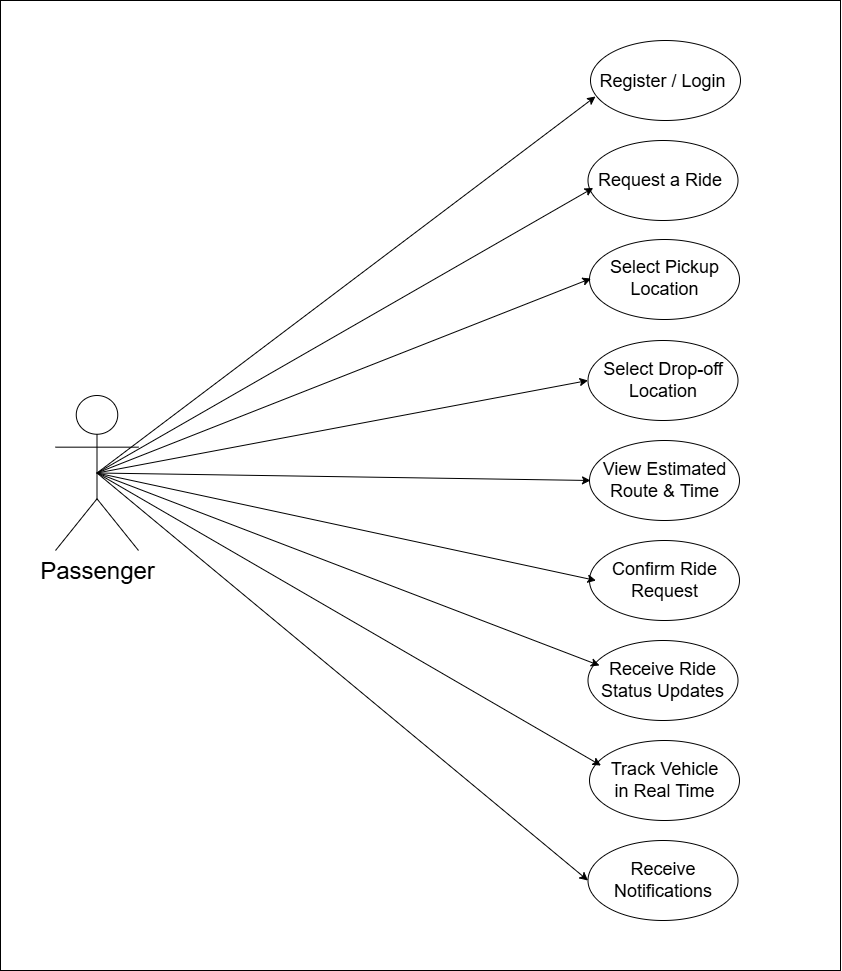}
\caption{ Passenger Use case diagram }
\label{fig:use_case}
\end{figure}

GO-DRiVeS allows users to preform multiple functionalities some of them are mentioned in the use case diagram in figure \ref{fig:use_case}.


\begin{figure}[h]
\centering
\includegraphics[width=0.9\textwidth]{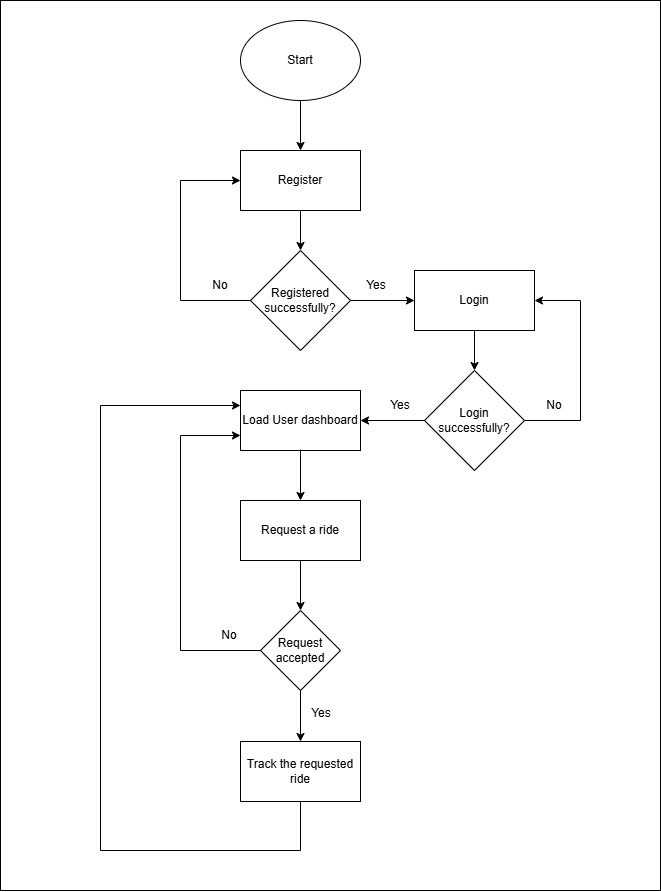}
\caption{GO-DRiVeS System Flow }
\label{fig:system-flow}
\end{figure}

As shown in figure \ref{fig:system-flow} this is the flow that GO-DRiVeS application followed.

\begin{figure}[h]
\centering
\includegraphics[width=0.9\textwidth]{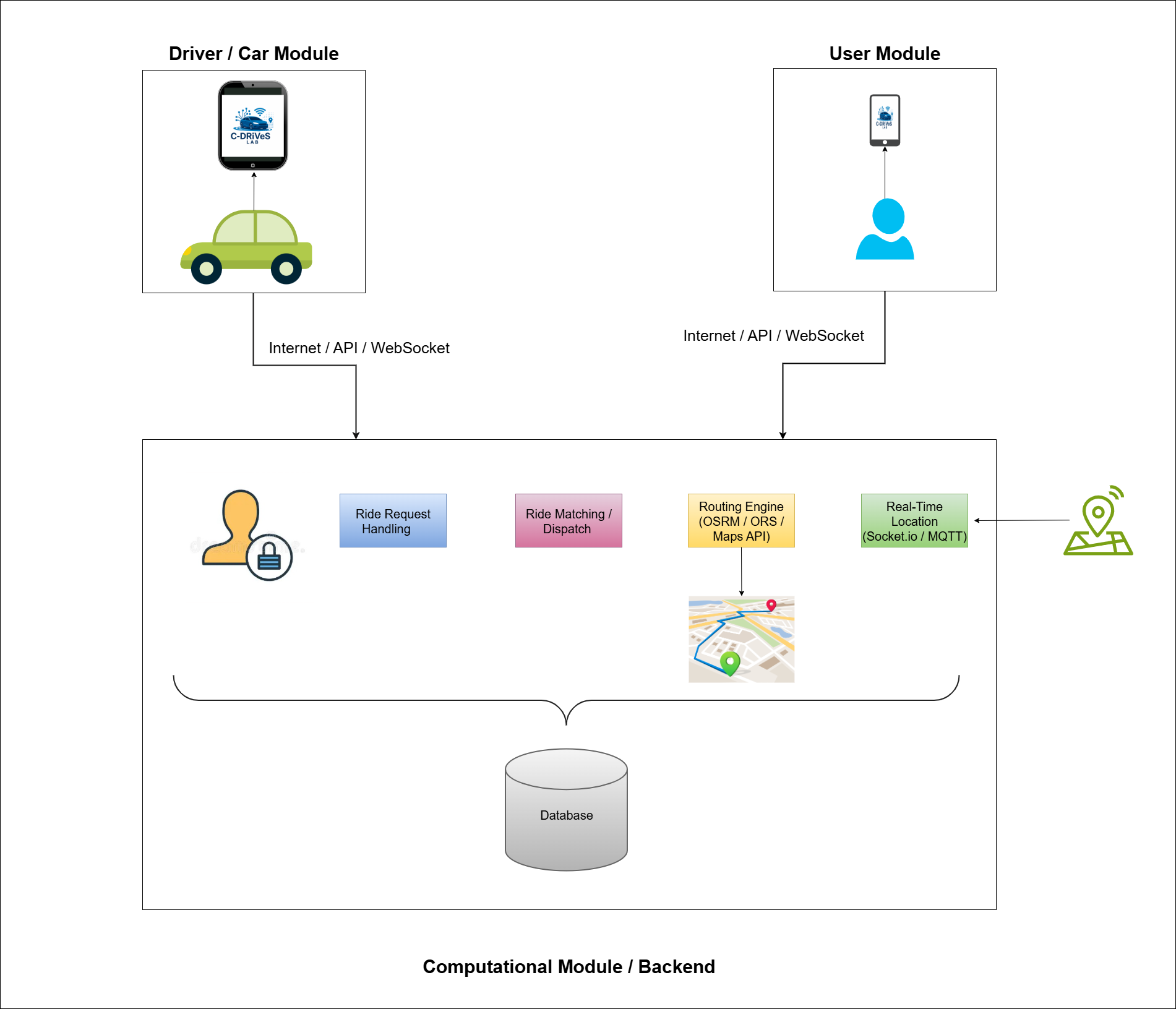}
\caption{GO-DRiVeS System Pipeline }
\label{fig:system-pipeline}
\end{figure}

GO-DRiVeS followed a specific pipeline when handling communications and interactions between the users, drivers, and the backend / server. illustrated in figure \ref{fig:system-pipeline}. 

\subsection{Tools and Technology}
This section outlines the tools, libraries, and frameworks that were used in the development of the GO-DRiVeS application, along with the reason behind choosing them.

\begin{itemize}
    \item \textbf{Frontend:} React native (Expo) was chosen to be used; as according to the comparison made between the frontend framework used in some papers' applications, as well as according to the most used frameworks in 2025; React native was from the top 2 options to choose from with flutter. But according to this paper \cite{stender2020cross}, which compares between react native and flutter. It was mentioned that for applications that need complex functionalities, React native is the best to use.

     \item \textbf{Backend:} Node.js and Extab were chosen to be used; as according to the comparison made between the backend framework used in some papers' applications , as well as according to the most used frameworks in 2025; Node.js was from the top 2 options to choose from with Firebase. But according to the libraries that node.js provides which nearly have everything, the decision was made that Node.js and extab are the ones to be used.

     \item \textbf{Database:} MongoDB was chosen to be used; as it was found that the type of database that suits the application is NoSQL according to the requirements. According to the most used NoSQL databases in 2024, MongoDB was from the top 2 options to choose from with firebase. But according to MongoDB's flexible schema and its horizontal scaling; it was decided that it was the better option to be used.

     \item \textbf{Communication:} Socket.IO was chosen to be used; as it was found that it is from the best communication libraries that are compatible to work with React native and Node.js, in addition to it being known to be best for real time events, location and chats.\\ Socket.Io is a JavaScript library for real time, that allows bidirectional, event based communication between clients and servers. The server is usually integrated into the backend, while the client is used in the frontend. As Socket.io is bidirectional communication, sometimes the client is the one sending the events / notifications and the server listens to them and other times the opposite.

     \item \textbf{Map:} Expo map was chosen to be used; as it integrates Google maps in it, in addition to it being provided to be used when using Expo, as well as not having to use an API to render the map.

    \item \textbf{Routing:} The free trial of ORS API was chosen to be used; as it provides the option to get the optimized version of the route without having to integrate any optimization techniques such as A* and Dijkstra. However, since the used type of ORS is built on top of GraphHopper,which is a high-performance routing engine, that uses Dijkstra’s algorithm, A* or Contraction Hierarchies to optimize its routes. Hence ORS uses these algorithms as well, but since this is the free trail, the optimization comes with some restrictions.

    \item \textbf{Tracking:} Expo location was chosen to be used; as it was provided  to be used with Expo, in addition to being able to get the GPS location.
\end{itemize}


\subsection{The User Perspective}
In GO-DRiVeS application, the user has the ability to interact with the application and go through all its steps starting with the :

\subsubsection{User Registration and Login}
This feature was the very first feature to be designed and implemented, as it is the first thing that the user encounter when wanting to use the GO-DRiVeS application. 

\paragraph{Registration process}
In this process the user whether it was a student or a faculty staff is able to create an account by entering their credentials; which are the users' university ID and mail , first and last names, phone number and last but not least, a password, which must be at least 6 numbers, letters or characters.

\paragraph{Validation Process}
This process happens right after the registration, as when the user finishes a pending page is shown to him as in the meantime those credentials that were entered are processed to the admin who by his role will check them, and if they were found valid, the admin will accept the user and an email will be sent to him with the acceptance and the ability to login. On the other hand, if the credentials were found to be invalid, then the admin will reject the user and an email will be sent to him stating that the credentials are not valid and to try to register again.

\paragraph{Login Process}
This process will be granted after the acceptance of the user. In it, the user will have to enter his username and password. The password will be the one that was entered in the registration process, however, the username will be auto generated for the user, and it will be the combination of the first and last names separated by a (.).
\subsubsection{Ride Request}
This feature is the main functionality of the GO-DRiVeS application, as it allows users to request on demand rides from their selected pickup location and destination within the GUC campus.
\paragraph{Where to process}

This process starts first in the User Dashboard, as the user has to tab on the "where to" button to be able to go to the next page. 

In the next page, users are given two options to choose from, either to pin their drop-off location themselves on the map or to search for the place, and it will be automatically pinned for them. After that, they have to tab on the confirm button and a notification will appear that confirms the desired drop-off location. Once the user tab on the OK button in the confirmation notification the longitude and latitude of the pinned drop-off location will be saved in the database.

\paragraph{From process}
This process is the same as the Where to process, however, users got an additional option to choose from which is to select their current location and it will be automatically pinned for them. In this process, the pinned location is saved in the pickup field instead of the drop-off. The rest of the steps are the same as in the Where to process . A notification will be sent to the user to confirm the pickup location that will be saved in the database .

\paragraph{Seats selection and confirmation processes}
In this process users have to enter the number of seats that they need for this ride, after that they have to tab on the "save number of seats" button. Once it is tabbed, a notification will appear confirming the number of needed seats, once the user tab on OK the number of seats requested will be saved in the database.

After setting the number of seats and in the same page, the pickup and drop-off locations that the user entered before will be present, in addition to the estimated distance and time that the ride between them will take. Users can also see the estimated route that will be taken from the pickup point to the drop-off one; through the map that has the pickup pin, drop-off pin and the route between them on it. Having verified everything, users have to tab on the "Confirm request" button, which will send the request to the car for acceptance or rejection.

\paragraph{Request process}
\begin{itemize}
    \item \textbf{From User and Car POV:} After the user confirmed the request and tabbed on the button, a notification appears to him that confirm that the request was sent to an available car and is waiting to be accepted . Once the user tabes on OK, he gets navigated to a waiting page until the request is accepted. When the driver in the car gets notified that there is a ride request, he first checks the number of seats that are requested and the available ones in the car; if they are compatible, then he accepts the ride request, and a notification appears to the user to notify him that a car accepted his request and to tab OK to be directed to the tracking screen. However, if the requested seats are not compatible with the available; ones that the driver rejects the request, and a notification appears to the user stating that the request was rejected and to try again later. Once the user tabes OK, he gets navigated back to the dashboard.
    \item \textbf{Logic behind: }When the user sends the request to the car; in the backend this happens through a real time communication library, which is Socket.IO . As mentioned in the Tools and Technology Selection and Justification, there are clients and servers in socket.io and either of them can emit the event and the other listen to it. Here in sending the ride request some steps were followed to achieve that: 
        \begin{enumerate}
             \item Once the user tabes on the confirmation button, it gets translated into a post request with the pickup, drop-off locations and number of seats to a backend method called "Confirm-ride".
             
            \item In the Confirm-ride method, it searches in the database for the available cars, then sends / emits a notification to each of them containing all the request details.
            
            \item The listener for this emitted notification is in the "handle-accept-ride" method in the car dashboard, which extracts the details of the request from the received notification and posts them to a backend method called "accept-ride", which checks everything related to the car,then send back an accept message with the type "ride-accepted". In addition to sending the user, a notification that the ride was accepted based on the retrieved message from the "accept-ride" backend method.

            \item The listener for the accept notification is in the waiting page, which listens for the type / status of the ride that is sent with the notification which is "ride-accepted". Then it shows the user the notification that his ride was accepted and to tab OK to navigate to the tracking page.

        \end{enumerate}

    \end{itemize}


\subsubsection{Tracking}

, This section is the second main functionality in the GO-DRiVeS application, as it allows users to know when and from where the car is coming.\\ This section includes two primary sections: real time tracking and routing.

\paragraph{Routing}
Routing is considered to be one of the most important features that any transportation application must have. As without it no other function can be completed; for instance, it plays an important role in the tracking process.

Routes can be implemented using various tools, some of them are manual while others are already implemented and ready to be used, such as the one chosen to be used in the GO-DRiVeS application which is ORS. As mentioned previously ORS APIs give their users the option to optimize their routes, and that is what was used to make all the routes in the application optimized.

In the application there are some stages that the route goes through: Start Journey, Head to Pickup, I Have Arrived, Start Ride, End Ride and Finish

These stages are all used by the car , while only two of them are used by the user "I Have Arrived" and "End Ride". In addition to these stages, there is route status: "enroute", "arrived"  and  "completed".

\paragraph{Tracking}
As mentioned above, there are some stages and status for the route, those stages and status are used to ease the tracking process for both of the user and the driver. And that is illustrated in the following steps:

\begin{enumerate}
    \item After the driver accepted the ride request, he is navigated to his track screen which initially is in the start journey stage, then the driver manually tabes on a button that changes the ride stage to "head to pickup".
    \item Once in the "head to pickup stage" the ride's pickup statues is set to be "enroute".In addition to making a route that connects between the current location of the car using the gps and the pickup point.At the same time, the user can also see the same route; as the gps location of the car is saved in the database by the car, and as the application is unified the gps location can be reached by the user.
    
    The effect of seeing the car moving in both of the driver and the user sides is done through retrieving the gps location from the database every amount of time and making a route using this location.
    \item Once the driver arrives at the pickup point, he tabes on a button that sets the ride stage to "i have arrived",as well as setting the ride status to "arrived". In addition to emitting a notification to the user side stating that the driver has arrived and setting the type to "driver-arrived".
    
    At the user side there is a listener that listens on the type, once it receives the type "driver-arrived" it shows the user a notification stating that the driver has arrived at the pickup point.
    
    Then a route between the current location of the car and the drop-off location is made on both the user and driver's sides.
    
    \item Once the user gets on board, the driver tabes on the button, which changes the ride stage to "Star ride",as well as setting the ride's pickup status to "compleat" and the drop-off status to "enroute". After that, the tracking starts again and the users are able to see the car moving on the map.
    
    \item Once the driver reaches the drop-off location, he tabes on the button which sets the route stage to "End route", as well as setting the drop-off status to "completed". In addition to emitting a notification to the user side stating that the driver has arrived and setting the type to "ride-ended".
    
    At the user side there is a listener that listens on the type, once it receives the type "ride-ended" it shows the user a notification stating that the driver has arrived at the drop-off point.
\end{enumerate}

Once the ride finishes, on the user side, the user is navigated back to the dashboard and can view all the notifications that were received along the ride. On the other hand, the ride's stage gets automatically changed to "finished", and then navigates to the car dashboard.
\section{Results}\label{sec4}

In this section, multiple experiments have been conducted, and an experiment full ride request will be demonstrated through some sample videos of the full ride showing each feature in the application, along side the full ride video.

Registration and login process:
This process validates the authentication and session management workflow of GO-DRiVeS. As the system correctly enforced users to enter correct and authorized credentials before allowing access to any ride related functionalities. By that, the system is confirmed to preserve users' data integrity and ensure consistent user identity management across multiple sessions.

\par This is a video sample of an experiment conducted; the user is trying to register to the application by entering all the credentials, then after gaining permission to login, the user logs in with the user name and password. The experiment can be illustrated using the following link: 


\nolinkurl{https://drive.google.com/file/d/1NjaNOX8q89QeBYaeXh_JqxLg0kjz571I/view?usp=sharing}

Ride requesting process: 

This process is the request creation pipeline, as the user specifies a dropoff location, pickup point, and the required seat number. With that, the system successfully generates a ride request entity, and the request that is ready for dispatch. This process is seen in this experiment video sample, the user is requesting a ride by first entering / pinning the drop-off location, then choosing the pickup point, followed by selecting the number of seats needed in this request. Then the user checks the trip details, and lastly confirms the request. After that, the user waits for some time for the trip to get accepted, and once it gets accepted the user is directed to the tracking pages. The experiment can be illustrated using the following link:


\nolinkurl{https://drive.google.com/file/d/1N_hxw5LmRaBazK59rMS5YCMlD7W1KIgH/view?usp=sharing}
    
Pickup tracking process:

This process validates the real-time synchronization between the driver movement and the passenger interface. The passenger has the ability to observe and track the optimized route used by the driver to reach the pickup point, in addition to receiving arrival notifications once the vehicle reaches the designated pickup location.
\par In the experiment video sample, the user is shown the optimized route the driver has taken to reach the pickup point, and the arrival notification that was received . The experiment can be illustrated using the following link: 


\nolinkurl{https://drive.google.com/file/d/1N6TS9UQIsE1A1Bm_ASqZqcTyysePAc8Z/view?usp=sharing}
    
Dropoff tracking process: In this video sample the user can track the optimized route the car is taking, however, in this experiment the route that was chosen had a blockage in real life and the car could not go through; that is why the driver chose to go through a different route, despite that as seen in the video the route does not get changed directly when the driver chose the other way as the routing optimization API is still seeing that this is the most optimized route that must be taken to reach the destination, but once thee car reaches a specific point the route gets changed to the other optimized one. At the end, when the car reached the dropoff point the user got notified with that. The experiment can be illustrated using the link: 


\nolinkurl{https://drive.google.com/file/d/1N6m2mbj7xdDZoDKkSCY84zWQsaQ48Ltw/view?usp=sharing}
    
This is the complete ride experiment that was conducted to evaluate the system’s end to end stability, covering registration, ride requesting, pickup tracking, and drop-off tracking and confirmation. The experiment confirms that GO-DRiVeS successfully maintained the full ride cycle consistency.
Full ride experiment can be illustrated using the following link:

\nolinkurl{https://drive.google.com/file/d/1aWjHEtyrgMlkZqjJ-QTZJObygiJ6DTvY/view?usp=sharing}
\section{Conclusion and Future Recommendations}\label{sec4}

This paper introduces the GO-DRiVeS application, which is an on demand ride sharing and requesting app made specifically for university students and staff to save them from long walks around the campus to reach a specific place. 

This paper Firstly, start with an introduction of the history, background, and evolution of transportation hailing to transportation applications in a timeline manner. 

Followed by, a review summary and comparisons made between the already documented transportation applications, these comparisons mainly focus on comparing the frameworks that these applications used. In addition to extracting the primary and secondary functionality that any transportation application has to have. Lastly, an overview of the types of system architecture used in these applications.

Followed by, the phases that this paper application went through. Starting with the chosen project management methodology which is Agile and the reason behind choosing it. Then stating the functional requirements that allow the users to interact with the application through. Followed by, the system pipeline that was used based on the reviewed papers and external research. 

Afterwords, the system flow is shown in a big picture manner , followed by its detailed illustration.

Then the results that were conducted by some experimental trials and a sample of them was demonstrated that confirmed the system's stable behavior towards handling ride requests and completing them efficiently.

Although the GO-DRiVeS application's system successfully implemented its essential ride-request features, in addition to including FIFO (First-In, First-Out) request handling and real-time vehicle tracking, however there are some functionalities that need to be worked on more.

\begin{itemize}
    
    \item Improved Multi-Request Management. 
    \item  Route Optimization with VROOM and ORS
    \item Deployment.
    
\end{itemize}

\appendices

\bibliographystyle{IEEEtran}
\bibliography{sections/ref} 

\end{document}